\documentclass[12pt,doublespacing]{article}

\usepackage{fullpage}
\usepackage{setspace}
\usepackage{authblk}

\usepackage{cite}
\usepackage{hyperref} 

\usepackage{graphicx}

\usepackage{amsmath}
\usepackage{amssymb}
\allowdisplaybreaks

\usepackage{float}
\usepackage{subfig}
\usepackage{makecell}
\usepackage{multirow}
\usepackage{rotating}

\usepackage{enumitem}

\usepackage{algorithmic}
\newcounter{manuscriptalgorithm}

\begin{document}

\title{A Contrast-Source Inversion Scheme Based on Stochastic Optimization and Plug-and-Play Regularization}

\author[]{Lingqi Gao}
\author[]{Hakan Bagci\vspace{0.5cm}}

\affil[]{Electrical and Computer Engineering (ECE) Program
\authorcr Computer, Electrical, and Mathematical Science and Engineering (CEMSE) Division
\authorcr King Abdullah University of Science and Technology (KAUST)
\authorcr Thuwal, 23955-6900, Saudi Arabia
\authorcr e-mail: lingqi.gao@kaust.edu.sa}

\date{}
\maketitle
\newpage

\begin{abstract}
An electromagnetic inversion scheme that integrates stochastic optimization (STO) and plug-and-play (PNP) regularization into contrast-source inversion (CSI), termed STO-PNP-CSI, is developed. Standard CSI solves for the contrast source vector of every transmitter at each iteration, which is expensive in a multi-transmitter configuration. STO instead solves for only one randomly selected contrast source vector per iteration, which reduces the per-iteration cost and can help the inversion escape poor local minima and saddle points. The resulting loss of information, however, increases the ill-posedness of the inversion. To counter this, the Swin-Conv-UNet (SCUNet) denoiser is plugged into the CSI scheme as an implicit regularizer, supplying a learned prior that is stronger than conventional hand-crafted ones and stabilizes the reconstruction. The proposed STO-PNP-CSI is applied to both synthetic and experimental data. The results show that it yields accurate reconstructions at substantially lower computational cost than CSI, including under strong nonlinearity and measurement noise.
\end{abstract}

\section{Introduction}\label{introduction}

Electromagnetic (EM) inversion aims to estimate the spatial distribution of electrical parameters such as permittivity and conductivity of an investigation domain from external measurements~\cite{Microwave Imaging}. Because the measurements are taken externally, the technique is noninvasive, and the required instrumentation is inexpensive compared with many alternative imaging methods. For these reasons, it has been widely used in various fields such as biomedical imaging~\cite{Stroke diagnosis using microwave techniques: Review of systems and algorithms}, geophysical exploration~\cite{Ground Penetrating Radar}, infrastructure assessment~\cite{Condition assessment of the buried utility service infrastructure}, and nondestructive testing~\cite{Electromagnetic non-destructive evaluation techniques for the monitoring of water and chloride ingress into concrete: a comparative study}.

Over the past few decades, EM inversion algorithms have evolved from linear inversion based on the Born approximation~\cite{Microwave Imaging} to iterative schemes that account for the nonlinearity of the inverse scattering problem. The latter include the Born iterative method (BIM)~\cite{Microwave imaging for brain stroke detection using Born iterative method}, the distorted Born iterative method (DBIM)~\cite{Reconstruction of two-dimensional permittivity distribution using the distorted Born iterative method} and related Newton-type schemes~\cite{Microwave imaging-complex permittivity reconstruction with a Levenberg--Marquardt method, Nonlinear microwave imaging for breast-cancer screening using Gauss--Newton method and the CGLS inversion algorithm}, contrast-source inversion (CSI)~\cite{A contrast source inversion method}, and general optimization methods such as conjugate gradient (CG)~\cite{Conjugate gradient method applied to inverse scattering problem}.

Among these methods, CSI has been proven to be an effective and robust algorithm. The fundamental idea underlying CSI is to update the contrast and the contrast source alternately, which splits the nonlinear EM inversion into two linear subproblems~\cite{A contrast source inversion method}. CSI has been applied to a range of EM inversion tasks, including free-space inverse scattering~\cite{Application of the multiplicative regularized contrast source inversion method on TM-and TE-polarized experimental Fresnel data}, geo-EM inversion~\cite{An integral equation approach for 2.5-dimensional forward and inverse electromagnetic scattering}, joint inversion~\cite{Study on joint inversion algorithm of acoustic and electromagnetic data in biomedical imaging}, and electrical property tomography~\cite{CSI-EPT: A contrast source inversion approach for improved MRI-based electric properties tomography}. Building on the standard CSI framework, several variants have been developed to improve its performance, such as the extended CSI~\cite{Extended contrast source inversion}, the cross-correlated CSI~\cite{Cross-correlated contrast source inversion}, and the wavelet-domain CSI~\cite{A contrast source inversion method in the wavelet domain}. In this work, standard CSI is made more efficient and robust by integrating stochastic optimization (STO) and plug-and-play (PNP) regularization. The resulting method is termed STO-PNP-CSI.

STO traces back to the stochastic gradient descent (SGD) algorithm~\cite{Stochastic gradient descent tricks} and has since been extended in various forms~\cite{Adam: A method for stochastic optimization}. Today, STO is widely used to train deep neural networks. The fundamental principle underlying STO is to optimize using random subsets of the data, which significantly reduces the computational cost of large-scale optimization problems. In addition, the randomness in these methods introduces stochastic perturbations that can help the optimization escape poor local minima and saddle points~\cite{Escaping from saddle points-online stochastic gradient for tensor decomposition}. Note that methods such as Bayesian inversion and global optimization are, in many contexts, also referred to as STO. In this paper, STO refers only to the concept defined in this paragraph.

In this work, the STO concept is applied to CSI. Unlike SGD, which randomizes over data samples to form a stochastic gradient, the proposed method solves for a single, randomly selected contrast source vector at each iteration and uses it to update the contrast. This is motivated by the observation that computing all contrast source vectors at each iteration is time-consuming, especially when the number of transmitters is large, so updating only one reduces the per-iteration cost. Moreover, as noted above, the stochastic updates can help CSI escape poor local minima and saddle points. This capability is well suited to EM inversion, where the nonlinearity strengthens as the contrast and operating frequency increase~\cite{Degree of nonlinearity and a new solution procedure in scalar two-dimensional inverse scattering problems, A new integral equation method to solve highly nonlinear inverse scattering problems} and traditional algorithms often fail to reach the global minimum. The STO strategy therefore improves the ability of CSI to handle highly nonlinear problems.

However, the stochastic updates also increase the ill-posedness of the inversion. Because only a single transmitter is used at each iteration, information is lost relative to standard CSI, which requires a strong regularizer for convergence~\cite{Parameter Estimation and Inverse Problems}. Regularization has been studied extensively in EM inversion~\cite{Overview and classification of some regularization techniques for the Gauss--Newton inversion method applied to inverse scattering problems}. Conventional methods add a regularization term to the data fidelity term and optimize them together. Representative examples include Tikhonov regularization, which is defined by the square of the $\ell_2$-norm of the model~\cite{Tikhonov regularization and total least squares}; sparsity-promoting regularization, which is defined by the $\ell_0$-norm, $\ell_1$-norm, or $\ell_1/\ell_2$-norm of the model or its gradient~\cite{Shrinkage-thresholding enhanced Born iterative method for solving 2D inverse electromagnetic scattering problem, A preconditioned inexact Newton method for nonlinear sparse electromagnetic imaging, Non-linear inverse scattering via sparsity regularized contrast source inversion, A ratio-norm regularization for sparsity-promoting electromagnetic inversion, A Sparsity-Promoting Electromagnetic Inversion Method Regularized by L1-L2 Norm of the Model Gradient}; and multiplicative regularization~\cite{Extended contrast source inversion}. Each of these methods encodes prior information through a fixed, closed-form expression. Such hand-crafted priors, however, are often too simple to capture the complexity of realistic profiles. Consequently, conventional regularization is unlikely to yield satisfactory reconstructions in CSI that integrates an STO strategy, which is more ill-posed than standard CSI.

In this work, to supply this regularization, a PNP scheme is integrated into the STO-based CSI scheme. PNP regularization uses a proximal algorithm to split the optimization into a data fidelity update and a regularization update. The regularization update can be interpreted as an image denoising problem~\cite{Plug-and-play priors for model based reconstruction}. This allows an implicit prior to be imposed through a denoiser, so that any state-of-the-art denoiser can be used in place of an explicit regularizer. PNP regularization was at first developed in image processing~\cite{Plug-and-play priors for model based reconstruction} and later adopted for nonlinear inverse scattering, using block-matching and 3-D filtering (BM3D) as the denoiser~\cite{Image Denoising by Sparse 3-D Transform-Domain Collaborative Filtering, A plug-and-play priors approach for solving nonlinear imaging inverse problems}. More recently, advances in deep learning have produced neural-network denoisers trained on large, high-quality image datasets. Having learned realistic image priors, they serve as effective PNP denoisers and improve reconstruction accuracy~\cite{Plug-and-play image restoration with deep denoiser prior, An electromagnetic inversion scheme regularized by deep plug-and-play denoiser}. In this work, the recently proposed SCUNet denoiser~\cite{Practical blind image denoising via Swin-Conv-UNet and data synthesis} is adopted as the PNP regularizer for the strong denoising performance afforded by its novel network design. Note that a preliminary version of the resulting STO-PNP-CSI is reported in~\cite{Contrast-Source Inversion with Stochastic Optimization and Plug-and-Play Regularization}.

The rest of this paper is organized as follows. Section~\ref{formulation} first presents the standard CSI scheme and then develops the proposed STO-PNP-CSI scheme. Section~\ref{numerical_examples} reports numerical examples for both synthetic and experimental data, analyzing the performance of STO-PNP-CSI across different system configurations and levels of nonlinearity and noise, and comparing it with other methods. Section~\ref{conclusion} concludes the paper.

\section{Formulation}\label{formulation}

\subsection{Scattering Equations and Discretization}\label{scattering_equations}

\begin{figure}[ht]
\centering
\includegraphics[width=0.5\columnwidth]{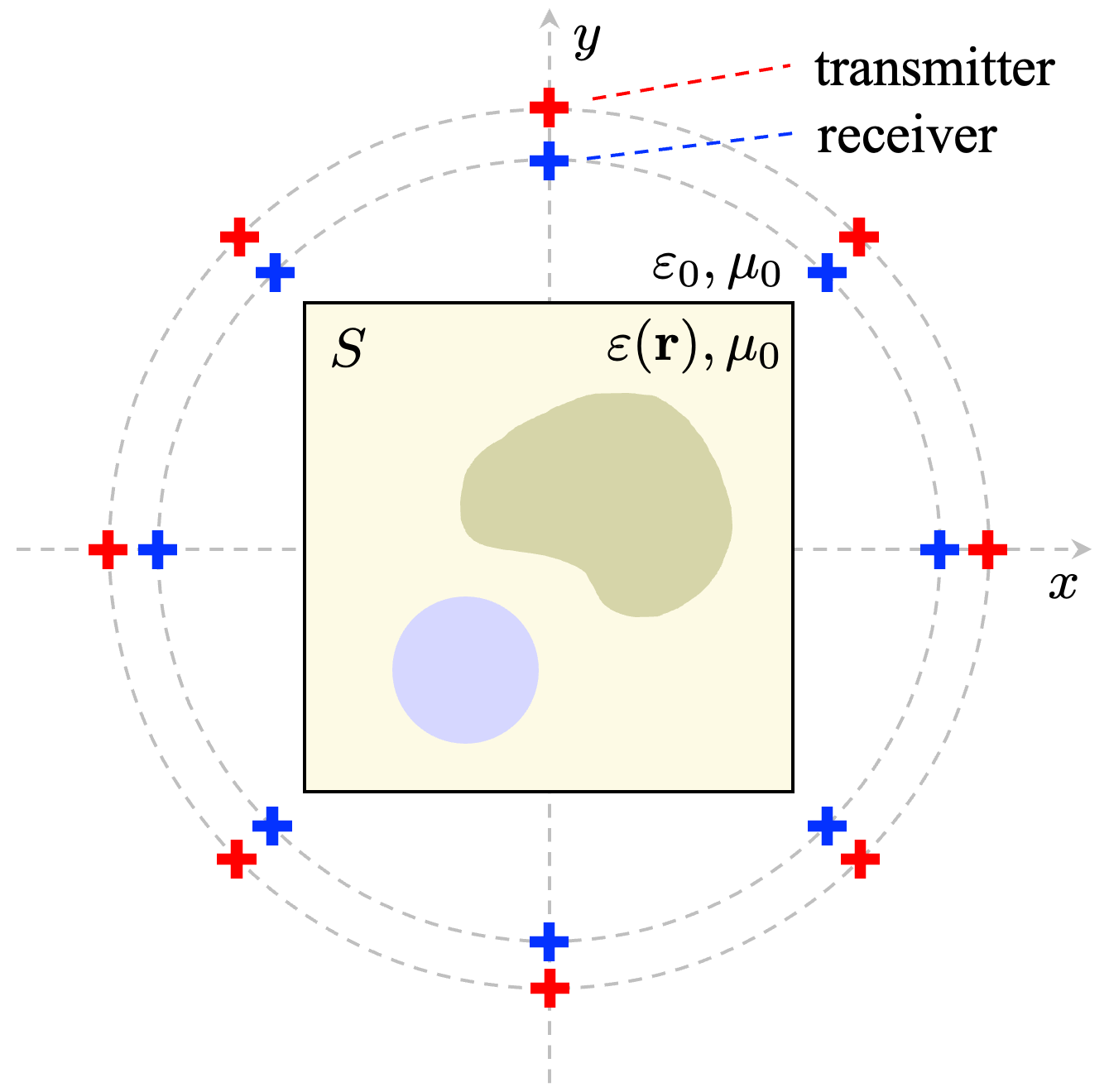}
\caption{Configuration of the inverse scattering problem.}\label{intro_configuration}
\end{figure}

The configuration shown in Fig.~\ref{intro_configuration} is used for EM inversion. An investigation domain $S$ with an inhomogeneous permittivity distribution $\varepsilon(\mathbf{r})$ is embedded in an unbounded homogeneous background medium with permittivity $\varepsilon_0$. The permeability in $S$ and in the background medium is $\mu_0$. The investigation domain is assumed to be lossless, so that $\varepsilon(\mathbf{r})$ is real. The contrast is defined as
\begin{equation}
\tau(\mathbf{r})= \begin{cases}\varepsilon(\mathbf{r}) / \varepsilon_0-1, & \mathbf{r} \in S \\ 0, & \mathrm{otherwise}\end{cases}.
\label{contrast_definition}
\end{equation}
To retrieve $\tau(\mathbf{r})$, $N_{\mathrm{t}}$ line-source transmitters, indexed by $i$, $i=1,2,\ldots, N_{\mathrm{t}}$, are located around $S$ and illuminate it one at a time. The scattered fields are measured by $N_{\mathrm{r}}$ receivers. A time-harmonic dependence $e^{\mathrm{j}\omega t}$, with $\omega$ the angular frequency, is assumed and suppressed. The fields are assumed invariant along the $z$-axis, reducing the problem to two dimensions (2-D) in the $xy$-plane. Under the transverse magnetic (TM) polarization, the electric field has only a $z$-component. At a given frequency, for transmitter $i$, the total field satisfies
\begin{equation}
E_i^{\mathrm{tot}}(\mathbf{r})=E_i^{\mathrm{inc}}(\mathbf{r})+E_i^{\mathrm{sca}}(\mathbf{r})\label{total_field}
\end{equation}
where $E_i^{\mathrm{inc}}(\mathbf{r})$, $E_i^{\mathrm{tot}}(\mathbf{r})$, and $E_i^{\mathrm{sca}}(\mathbf{r})$ are the $z$-components of the incident, total, and scattered fields, respectively. The scattered field is expressed as
\begin{equation}
E_i^{\mathrm{sca}}(\mathbf{r})=-k_0^2 \int_S \tau(\mathbf{r}^{\prime}) E_i^{\mathrm{tot}}(\mathbf{r}^{\prime}) g(\mathbf{r}, \mathbf{r}^{\prime})\,ds^{\prime}\label{scattered_field_integral}
\end{equation}
where $k_0=\omega \sqrt{\mu_0 \varepsilon_0}$ is the wavenumber of the background medium and $g(\mathbf{r}, \mathbf{r}^{\prime})=({\mathrm{j}}/{4}) H_0^{(2)}(k_0\left|\mathbf{r}-\mathbf{r}^{\prime}\right|)$ is the 2-D scalar Green function of the background medium, and $H_0^{(2)}(\cdot)$ is the zeroth-order Hankel function of the second kind. Substituting~\eqref{scattered_field_integral} into~\eqref{total_field} yields
\begin{equation}
E_{i}^{\mathrm{tot}}(\mathbf{r}) + k_0^2 \int_S \tau(\mathbf{r}^{\prime}) E_{i}^{\mathrm{tot}} (\mathbf{r}^{\prime}) g(\mathbf{r}, \mathbf{r}^{\prime})\,ds^{\prime}=E_{i}^{\mathrm{inc}}(\mathbf{r}).\label{state_integral_equation}
\end{equation}
To discretize~\eqref{scattered_field_integral} and~\eqref{state_integral_equation}, $S$ is divided into $N$ square cells $S_n$, $n = 1,2,\ldots,N$. Over these cells, $\tau(\mathbf{r})$ and $E_i^{\mathrm{tot}}(\mathbf{r})$ are expanded as
\begin{equation}
\begin{aligned}
\tau(\mathbf{r}) & =\sum_{n=1}^N\{\bar{\tau}\}_n p_n(\mathbf{r})\\
E_i^{\mathrm{tot}}(\mathbf{r}) & =\sum_{n=1}^N\{\bar{E}_i^{\mathrm{tot}}\}_n p_n(\mathbf{r})
\end{aligned}\label{field_contrast_expansion}
\end{equation}
where $\bar{\tau}$ and $\bar{E}_i^{\mathrm{tot}}$ store the samples of $\tau(\mathbf{r})$ and $E_i^{\mathrm{tot}}(\mathbf{r})$, with entries $\{\bar{\tau}\}_n = \tau(\mathbf{r}_n)$ and $\{\bar{E}_i^{\mathrm{tot}}\}_n = E_i^{\mathrm{tot}}(\mathbf{r}_n)$, and $\mathbf{r}_n$ denotes the center of cell $S_n$. The pulse basis function $p_n(\mathbf{r})$ is defined as
\begin{equation}
p_n(\mathbf{r})= \begin{cases}1, & \mathbf{r} \in S_n \\ 0, & \mathrm{otherwise}\end{cases}.
\label{pulse_basis}
\end{equation}
Substituting~\eqref{field_contrast_expansion} into~\eqref{state_integral_equation} and sampling the resulting equation at the cell centers $\mathbf{r}_m$, $m = 1,2,\ldots,N$, yields the linear equation system
\begin{equation}
[\bar{\bar{I}}+\bar{\bar{G}}\,\mathrm{diag}\{\bar{\tau}\}]\,\bar{E}^{\mathrm{tot}}_{i}=\bar{E}^{\mathrm{inc}}_{i}\label{discrete_state_equation}
\end{equation}
where $\bar{E}^{\mathrm{inc}}_{i}$ stores the samples of $E_i^{\mathrm{inc}}(\mathbf{r})$, with entries $\{\bar{E}^{\mathrm{inc}}_i\}_n = E_i^{\mathrm{inc}}(\mathbf{r}_n)$, $\bar{\bar{I}}$ is the identity matrix, and $\mathrm{diag}\{\cdot\}$ maps a vector to a diagonal matrix. The contrast source corresponding to transmitter $i$ is defined as 
\begin{equation}
\bar{w}_i = \mathrm{diag}\{\bar{\tau}\}\,\bar{E}^{\mathrm{tot}}_i.
\label{contrast_source}
\end{equation}
In~\eqref{discrete_state_equation}, the integrated Green function matrix $\bar{\bar{G}}$ relates the contrast source $\bar{w}_i$ to the scattered field at the cell centers, and its entries are given by
\begin{equation}
\{\bar{\bar{G}}\}_{mn}=k_0^2\int_{S_n} g(\mathbf{r}_m, \mathbf{r}^{\prime})\,ds^{\prime}, \ m,n = 1,2,\ldots,N.\label{green_matrix}
\end{equation}
Substituting~\eqref{field_contrast_expansion} into~\eqref{scattered_field_integral} and sampling the resulting equation at the receiver locations $\mathbf{r}_j^{\mathrm{r}}$, $j=1,2,\ldots,N_{\mathrm{r}}$, yields the linear equation system
\begin{equation}
\bar{E}^{\mathrm{sca}}_{i}=\bar{\bar{H}}\,\mathrm{diag}\{\bar{E}^{\mathrm{tot}}_{i}\}\,\bar{\tau}=\bar{\bar{H}}\,\mathrm{diag}\{\bar{\tau}\}\,\bar{E}^{\mathrm{tot}}_{i}
\label{discrete_data_equation}
\end{equation}
where $\bar{E}^{\mathrm{sca}}_{i}$ stores the samples of $E_i^{\mathrm{sca}}(\mathbf{r})$ at the receiver locations, with entries $\{\bar{E}^{\mathrm{sca}}_i\}_j = E_i^{\mathrm{sca}}(\mathbf{r}_j^{\mathrm{r}})$. The integrated Green function matrix $\bar{\bar{H}}$ relates the contrast source $\bar{w}_i$ to the scattered field at the receivers, and its entries are given by
\begin{equation}
\label{receiver_green_matrix}
\{\bar{\bar{H}}\}_{jn}=-k_0^2\int_{S_n} g(\mathbf{r}_j^{\mathrm{r}}, \mathbf{r}^{\prime})\,ds^{\prime}, \  j = 1,2,\ldots,N_{\mathrm{r}},\ n = 1,2,\ldots,N.
\end{equation}
For a given $\bar{\tau}$, $\bar{E}_i^{\mathrm{tot}}$ can be calculated by solving~\eqref{discrete_state_equation}. Then, $\bar{E}^{\mathrm{sca}}_{i}$ can be computed from~\eqref{discrete_data_equation}.

\subsection{CSI}\label{csi}

In CSI,~\eqref{discrete_state_equation} is referred to as the state equation and~\eqref{discrete_data_equation} as the data equation~\cite{A contrast source inversion method}. Using the contrast source $\bar{w}_i$ for transmitter $i$ defined in~\eqref{contrast_source} and letting $\bar{\bar{T}} = \mathrm{diag}\{\bar{\tau}\}$, they are rewritten as
\begin{equation}
\bar{\bar{T}}\bar{E}^{\mathrm{inc}}_i=\bar{w}_i+\bar{\bar{T}}\bar{\bar{G}}\bar{w}_i\label{csi_state_equation}
\end{equation}
\begin{equation}
\bar{E}^{\mathrm{sca}}_i = \bar{\bar{H}}\bar{w}_i.\label{csi_data_equation}
\end{equation}
The CSI objective function is constructed by enforcing~\eqref{csi_state_equation} and~\eqref{csi_data_equation} for all the transmitters. To balance the contributions of the two equations, each is normalized~\cite{A contrast source inversion method}. The resulting objective function reads
\begin{equation}
\Phi  =  \sum_{i=1}^{N_{\mathrm{t}}} \frac{1}{2}\frac{ \| \bar{\bar{T}}\bar{E}^{\mathrm{inc}}_i - \bar{w}_i - \bar{\bar{T}}\bar{\bar{G}}\bar{w}_i \|_2^2}{ \| \bar{\bar{T}}\bar{E}^{\mathrm{inc}}_i \|_2^2} + \sum_{i=1}^{N_{\mathrm{t}}} \frac{1}{2}\frac{ \| \bar{E}^{\mathrm{mea}}_i - \bar{\bar{H}}\bar{w}_i \|_2^2}{ \| \bar{E}^{\mathrm{mea}}_i \|_2^2}. 
\label{csi_objective_function}
\end{equation}
Here, $\bar{E}^{\mathrm{mea}}_i$ is the scattered field measured at the receivers in response to transmitter $i$. The first term of~\eqref{csi_objective_function} penalizes the residual of the state equation~\eqref{csi_state_equation}, and the second penalizes the residual of the data equation~\eqref{csi_data_equation}. 

The optimization of~\eqref{csi_objective_function} is carried out by alternately updating $\bar{w}_i$ and $\bar{\tau}$, leading to a two-step scheme. At iteration $k$, the two substeps read
\begin{equation}
\!\!\begin{aligned}
\bar{w}^{(k)}_i & = \arg\min_{\bar{w}_i} \bigg[\frac{1}{2}\frac{\| \bar{\bar{T}}^{(k-1)}\bar{E}^{\mathrm{inc}}_i - \bar{w}_i - \bar{\bar{T}}^{(k-1)}\bar{\bar{G}}\bar{w}_i \|_2^2}{\| \bar{\bar{T}}^{(k-1)}\bar{E}^{\mathrm{inc}}_i \|_2^2} \\
& + \frac{1}{2}\frac{ \| \bar{E}^{\mathrm{mea}}_i - \bar{\bar{H}}\bar{w}_i \|_2^2}{\| \bar{E}^{\mathrm{mea}}_i \|_2^2}\bigg], \ i = 1,2,\ldots,N_{\mathrm{t}}
\end{aligned}\label{csi_contrast_source_update}
\end{equation}
\begin{equation}
\bar{\tau}^{(k)} = \arg\min_{\bar{\tau}} \sum_{i=1}^{N_{\mathrm{t}}} \frac{1}{2} \| \mathrm{diag}\{\bar{\tau}\}\bar{E}^{\mathrm{tot},{(k)}}_i - \bar{w}^{(k)}_i \|_2^2.
\label{csi_contrast_update}
\end{equation}
Here, after~\eqref{csi_contrast_source_update} is solved, $\bar{E}^{\mathrm{inc}}_i - \bar{\bar{G}}\bar{w}^{(k)}_i$ in~\eqref{csi_contrast_update} is replaced by $\bar{E}^{\mathrm{tot},{(k)}}_i$ from~\eqref{csi_state_equation}.

\subsection{STO-PNP-CSI}\label{sto_pnp_csi}

The standard CSI scheme presented in Section~\ref{csi} is improved in two respects. The first is the STO strategy, which accelerates CSI and overcomes its limitations in handling strong nonlinearity. The second is the PNP regularization, which imposes a more informative prior to stabilize the inversion. The objective function of the proposed STO-PNP-CSI is written as
\begin{equation}
\Phi =  \sum_{i=1}^{N_{\mathrm{t}}} \frac{1}{2}\frac{\| \bar{\bar{T}}\bar{E}^{\mathrm{inc}}_i - \bar{w}_i - \bar{\bar{T}}\bar{\bar{G}}\bar{w}_i \|_2^2}{\| \bar{\bar{T}}\bar{E}^{\mathrm{inc}}_i \|_2^2} + \sum_{i=1}^{N_{\mathrm{t}}} \frac{1}{2} \frac{ \| \bar{E}^{\mathrm{mea}}_i - \bar{\bar{H}}\bar{w}_i \|_2^2}{ \| \bar{E}^{\mathrm{mea}}_i \|_2^2} + \lambda R(\bar{\tau})
\label{sto_pnp_objective_function}
\end{equation}
where $R(\bar{\tau})$ is the regularization term and $\lambda>0$ its weight.

To optimize~\eqref{sto_pnp_objective_function}, the alternating update of $\bar{w}_i$ and $\bar{\tau}$ is retained. In CSI,~\eqref{csi_contrast_source_update} is solved for each transmitter, so it must be solved $N_{\mathrm{t}}$ times per iteration to obtain $\bar{w}_i$ for every transmitter. This accounts for most of the computational cost, especially when $N_{\mathrm{t}}$ is large. To address this, inspired by the SGD algorithm and its variants~\cite{Stochastic gradient descent tricks, Adam: A method for stochastic optimization}, which optimize using a random subset of the data at each iteration, STO is applied to CSI. Unlike SGD, which randomizes over data samples to form a stochastic gradient, the proposed method randomly selects a single transmitter with index $p$ and solves for its contrast source $\bar{w}_p$, reducing the number of solves of~\eqref{csi_contrast_source_update} from $N_{\mathrm{t}}$ to one per iteration. Beyond this acceleration, the random selection introduces stochastic perturbations that can help the iteration move away from poor local minima and saddle points, improving convergence. This makes STO-PNP-CSI more effective at handling stronger nonlinearity caused by high operating frequency and large contrast. 

At iteration $k$, the two substeps for optimizing~\eqref{sto_pnp_objective_function} read 
\begin{equation}
\bar{w}^{(k)}_p = \arg\min_{\bar{w}_p} \bigg[\frac{1}{2}\frac{\| \bar{\bar{T}}^{(k-1)}\bar{E}^{\mathrm{inc}}_p - \bar{w}_p - \bar{\bar{T}}^{(k-1)}\bar{\bar{G}}\bar{w}_p \|_2^2}{\| \bar{\bar{T}}^{(k-1)}\bar{E}^{\mathrm{inc}}_p \|_2^2} + \frac{1}{2}\frac{ \| \bar{E}^{\mathrm{mea}}_p - \bar{\bar{H}}\bar{w}_p \|_2^2}{\| \bar{E}^{\mathrm{mea}}_p \|_2^2}\bigg]
\label{sto_contrast_source_update}
\end{equation}
\begin{equation}
\bar{\tau}^{(k)} = \arg\min_{\bar{\tau}} \frac{1}{2} \| \mathrm{diag}\{\bar{\tau}\}\bar{E}^{\mathrm{tot},(k)}_p - \bar{w}^{(k)}_p \|_2^2 + \lambda R(\bar{\tau}).\label{sto_contrast_update}
\end{equation}
Here, a fresh index $p$ is drawn at random from $\{1,2,\ldots,N_{\mathrm{t}}\}$ at each iteration. For simplicity,~\eqref{sto_contrast_source_update} is referred to as substep I, and~\eqref{sto_contrast_update} as substep II. 

Substep I is a linear least-squares problem, whose solution satisfies the normal equations
\begin{equation}
\bar{\bar{A}}^{(k)}_{\bar{w}_p} \bar{w}^{(k)}_p = \bar{b}^{(k)}_{\bar{w}_p}.\label{contrast_source_linear_system}
\end{equation}
Since $\bar{\tau}$ is real, $\bar{\bar{T}}^{(k-1)}$ is a real diagonal matrix and $\bar{\bar{T}}^{(k-1)\dagger} = \bar{\bar{T}}^{(k-1)}$, where ``$\dagger$'' denotes the conjugate transpose. The system matrix $\bar{\bar{A}}^{(k)}_{\bar{w}_p}$ and right-hand side $\bar{b}^{(k)}_{\bar{w}_p}$ are then given by
\begin{equation}
\bar{\bar{A}}^{(k)}_{\bar{w}_p} = \bar{\bar{H}}^{\dagger}\bar{\bar{H}} / \|\bar{E}^{\mathrm{mea}}_p\|_2^2 + \big(\bar{\bar{T}}^{(k-1)}\bar{\bar{G}} + \bar{\bar{G}}^{\dagger}\bar{\bar{T}}^{(k-1)} + \bar{\bar{G}}^{\dagger}\bar{\bar{T}}^{(k-1)}\bar{\bar{T}}^{(k-1)}\bar{\bar{G}} + \bar{\bar{I}}\big) / \|\bar{\bar{T}}^{(k-1)}\bar{E}^{\mathrm{inc}}_p\|_2^2 \label{contrast_source_system_matrix}
\end{equation}
\begin{equation}
\bar{b}^{(k)}_{\bar{w}_p}  = \bar{\bar{H}}^{\dagger}\bar{E}^{\mathrm{mea}}_p / \|\bar{E}^{\mathrm{mea}}_p\|_2^2 + \big(\bar{\bar{T}}^{(k-1)}\bar{E}^{\mathrm{inc}}_p  + \bar{\bar{G}}^{\dagger}\bar{\bar{T}}^{(k-1)}\bar{\bar{T}}^{(k-1)}\bar{E}^{\mathrm{inc}}_p\big) / \|\bar{\bar{T}}^{(k-1)}\bar{E}^{\mathrm{inc}}_p\|_2^2.
\label{contrast_source_rhs}
\end{equation}

After~\eqref{sto_contrast_source_update} is solved, as in CSI, $\bar{E}^{\mathrm{tot},{(k)}}_p = \bar{E}^{\mathrm{inc}}_p - \bar{\bar{G}}\bar{w}^{(k)}_p$ is used in~\eqref{sto_contrast_update} to update $\bar{\tau}$ using only $\bar{w}^{(k)}_p$. Conventionally, the regularization term $R(\bar{\tau})$ has a closed-form definition encoding the desired prior information about the solution. However, such hand-crafted terms often cannot represent an accurate prior. In PNP regularization, $R(\bar{\tau})$ need not be explicit. Instead, the associated update is carried out by a denoiser, which injects more realistic prior information. To this end, an  auxiliary variable $\bar{n}$ is introduced with the constraint $\bar{n}=\bar{\tau}$. Then, the optimization problem equivalent to~\eqref{sto_contrast_update} is
\begin{equation}
(\bar{\tau}^{(k)}, \bar{n}^{(k)}) = \arg\min_{\bar{\tau}, \bar{n}} \frac{1}{2} \| \mathrm{diag}\{\bar{\tau}\}\bar{E}^{\mathrm{tot},(k)}_p - \bar{w}^{(k)}_p \|_2^2+ \lambda R(\bar{n})\quad \mathrm{s.t.} \ \bar{n} = \bar{\tau}.\label{auxiliary_constrained_problem}
\end{equation}
Based on the half-quadratic splitting (HQS) algorithm~\cite{A new alternating minimization algorithm for total variation image reconstruction}, the penalty function of~\eqref{auxiliary_constrained_problem} is given by
\begin{equation}
L(\bar{\tau}, \bar{n}) = \frac{1}{2} \| \mathrm{diag}\{\bar{\tau}\}\bar{E}^{\mathrm{tot},(k)}_p - \bar{w}^{(k)}_p \|_2^2
+ \frac{\rho}{2}\|\bar{\tau} - \bar{n}\|_2^2 + \lambda R(\bar{n})
\label{hqs_penalty_function}
\end{equation}
where $\rho$ is the weight of the penalty term that enforces $\bar{n}=\bar{\tau}$ as a soft constraint. Equation ~\eqref{hqs_penalty_function} is then split into two substeps that alternately update $\bar{\tau}$ and $\bar{n}$
\begin{equation}
\bar{\tau}^{(k)} = \arg\min_{\bar{\tau}}\bigg[\frac{1}{2} \| \mathrm{diag}\{\bar{\tau}\}\bar{E}^{\mathrm{tot},(k)}_p - \bar{w}^{(k)}_p \|_2^2 + \frac{\rho}{2}\|\bar{\tau} - \bar{n}^{(k-1)}\|_2^2\bigg]\label{contrast_update_hqs}
\end{equation}
\begin{equation}
\bar{n}^{(k)} = \arg\min_{\bar{n}}\frac{1}{2(\sqrt{\lambda/\rho})^2}\|\bar{\tau}^{(k)} - \bar{n}\|_2^2 + R(\bar{n}).\label{denoiser_update_hqs}
\end{equation}
For simplicity,~\eqref{contrast_update_hqs} is referred to as substep II-a, and~\eqref{denoiser_update_hqs} as substep II-b.

Substep II-a is a linear least-squares problem, whose solution satisfies the normal equations
\begin{equation}
\bar{\bar{A}}^{(k)}_{\bar{\tau}} \bar{\tau}^{(k)} = \bar{b}_{\bar{\tau}}^{(k)}\label{contrast_linear_system}
\end{equation}
where $\bar{\bar{A}}^{(k)}_{\bar{\tau}}$ and $\bar{b}^{(k)}_{\bar{\tau}}$ are given by
\begin{equation}
\bar{\bar{A}}^{(k)}_{\bar{\tau}} = \mathrm{diag}\{\bar{E}_{p}^{\mathrm{tot},(k)}\}^{\dagger} \mathrm{diag}\{\bar{E}_{p}^{\mathrm{tot},(k)}\} + \rho \bar{\bar{I}}\label{contrast_system_matrix}
\end{equation}
\begin{equation}
\bar{b}^{(k)}_{\bar{\tau}} = \Re\Big\{\mathrm{diag}\{\bar{E}_{p}^{\mathrm{tot},(k)}\}^{\dagger} \bar{w}^{(k)}_{p}\Big\} + \rho \bar{n}^{(k-1)}\label{contrast_rhs}
\end{equation}
where $\Re\{\cdot\}$ returns the real part of its argument. The matrix $\bar{\bar{A}}^{(k)}_{\bar{\tau}}$ is real, since\\ $\mathrm{diag}\{\bar{E}_{p}^{\mathrm{tot},(k)}\}^{\dagger}\mathrm{diag}\{\bar{E}_{p}^{\mathrm{tot},(k)}\}$ is a real diagonal matrix; in $\bar{b}^{(k)}_{\bar{\tau}}$, the real part is taken because $\bar{\tau}$ is real while $\mathrm{diag}\{\bar{E}_{p}^{\mathrm{tot},(k)}\}^{\dagger}\bar{w}^{(k)}_{p}$ is complex.

From a statistical perspective, substep II-b in~\eqref{denoiser_update_hqs} can be interpreted as a nonblind Gaussian denoising problem, where $\bar{\tau}^{(k)}$ is the noisy image, $\bar{n}^{(k)}$ is the denoised image, and $\sigma=\sqrt{\lambda / \rho}$ is the noise standard deviation. This interpretation implies that, instead of solving the original optimization problem, one can directly apply a denoiser, thereby leveraging its implicit prior. As a result, any advanced denoiser, even without an explicit regularizer, can be incorporated into the inversion framework. The contrast map $\bar{\tau}$ is treated as a grayscale image, so an image denoiser can be applied directly.

\begin{sidewaysfigure}[ht]
\centering
\includegraphics[width=\columnwidth]{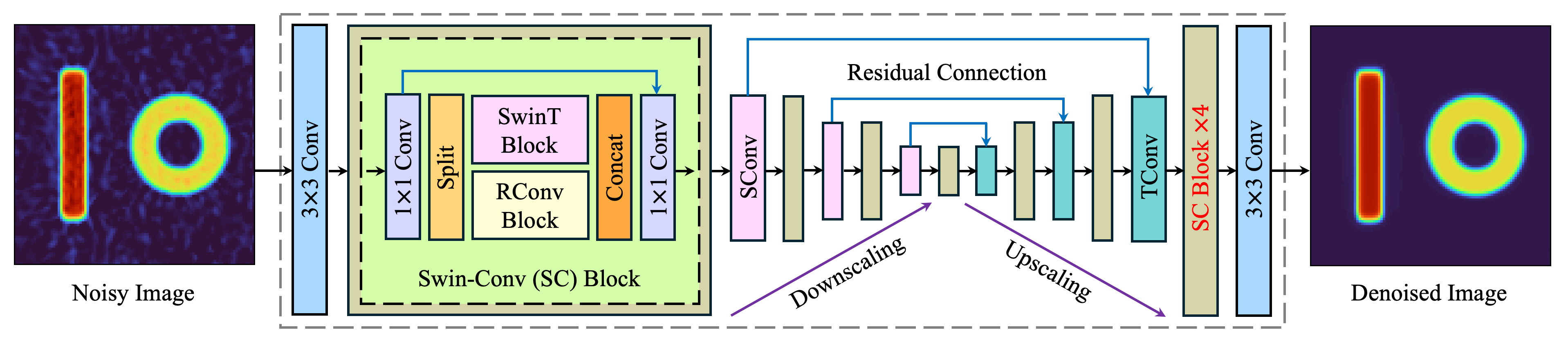}
\caption{Architecture of SCUNet~\cite{Practical blind image denoising via Swin-Conv-UNet and data synthesis}.}\label{scunet_architecture}
\end{sidewaysfigure}

To maximize the performance of PNP regularization, it is critical to select a proper denoiser. Denoisers based on supervised learning are especially preferred for two reasons. First, they are normally more effective, as they are trained on high-quality image-denoising datasets. Second, pretrained models can be used directly, making the denoising efficient. In this work, the recently proposed SCUNet is adopted as the PNP denoiser. As shown in Fig.~\ref{scunet_architecture}, SCUNet adopts a UNet architecture with four scales, using $64$, $128$, $256$, and $512$ channels, respectively. Downscaling and upscaling are performed by $2 \times 2$ strided convolution (SConv) and $2 \times 2$ transposed convolution (TConv) layers, respectively, connected by residual connections. In the SCUNet architecture, the Swin-Conv (SC) blocks fuse residual convolutional blocks and Swin Transformer blocks. In an SC block, the input tensor first passes through a $1 \times 1$ convolution and is then evenly split into two parts, which are processed by a Swin Transformer block and a residual convolutional block. Their outputs are concatenated and fed into another $1 \times 1$ convolution, with a residual connection linking the two $1 \times 1$ convolutions. This design enables SCUNet to capture both local and nonlocal image features and achieve denoising performance superior to that of existing neural-network denoisers.

During the inversion, substep I in~\eqref{sto_contrast_source_update}, substep II-a in~\eqref{contrast_update_hqs}, and substep II-b in~\eqref{denoiser_update_hqs} are executed alternately for a fixed number of iterations $N_{\mathrm{iter}}$. Substeps I and II-a amount to the linear systems~\eqref{contrast_source_linear_system} and~\eqref{contrast_linear_system}, which are solved by the generalized minimal residual (GMRES) algorithm~\cite{GMRES: A generalized minimal residual algorithm for solving nonsymmetric linear systems}. The matrix--vector products involving $\bar{\bar{G}}$ are accelerated by the fast Fourier transform (FFT) scheme~\cite{The CG-FFT Method: Application of Signal Processing Techniques to Electromagnetics}. Substep II-b is solved directly by applying SCUNet.

\subsection{Multi-Frequency Inversion}\label{multi_frequency}

The frequency-hopping scheme~\cite{A frequency-hopping approach for microwave imaging of large inhomogeneous bodies} is applied to the proposed STO-PNP-CSI method for multi-frequency inversion with $N_{\mathrm{f}}$ frequencies and $N_{\mathrm{iter}}$ iteration steps at each frequency. At the lowest frequency, the full objective function in~\eqref{sto_pnp_objective_function} cannot be optimized from a zero contrast, because for $\bar{\tau} = \bar{0}$ the state-term denominator $\|\bar{\bar{T}}\bar{E}^{\mathrm{inc}}_p\|_2^2$ vanishes~\cite{A contrast source inversion method}. The initial guess $\bar{\tau}^{(0)}$ is therefore obtained by first optimizing the data equation alone. At each subsequent frequency, the initial guess is the inversion result obtained at the previous frequency. After the inversion is completed at the end of the last frequency update, a final denoising step is applied to produce the result. Algorithm~\ref{sto_pnp_csi_algorithm} summarizes the workflow of STO-PNP-CSI.

\begin{figure}[!t]
\hrule\vspace{2pt}
\refstepcounter{manuscriptalgorithm}\label{sto_pnp_csi_algorithm}
\noindent\textbf{Algorithm \themanuscriptalgorithm} STO-PNP-CSI\par
\vspace{2pt}\hrule\vspace{2pt}
\footnotesize
\noindent \, 1: \textbf{for} $f = 1,2,\ldots,N_{\mathrm{f}}$\par
\vspace{1pt}
\noindent \, 2: \quad \textbf{if} $f = 1$\par
\vspace{1pt}
\noindent \, 3: \quad \quad obtain $\bar{\tau}^{(0)}$ by optimizing the data equation\par
\vspace{1pt}
\noindent \, 4: \quad \textbf{else}\par
\vspace{1pt}
\noindent \, 5: \quad \quad $\bar{\tau}^{(0)}$ is the result at frequency $f-1$\par
\vspace{1pt}
\noindent \, 6: \quad \textbf{end if}\par
\vspace{1pt}
\noindent \, 7: \quad initialize $\bar{n}^{(0)} = \bar{\tau}^{(0)}$\par
\vspace{1pt}
\noindent \, 8: \quad calculate $\bar{\bar{G}}$, $\bar{\bar{H}}$, and $\bar{E}^{\mathrm{inc}}_{i}$, $i=1,2,\ldots,N_{\mathrm{t}}$\par
\vspace{1pt}
\noindent \, 9: \quad \textbf{for} $k = 1,2,\ldots,N_{\mathrm{iter}}$\par
\vspace{1pt}
\noindent 10: \quad \quad generate random integer $p \in \{1,2,\ldots,N_{\mathrm{t}}\}$\par
\vspace{1pt}
\noindent 11: \quad \quad calculate $\bar{w}_{p}^{(k)}$ by solving~\eqref{contrast_source_linear_system}\par
\vspace{1pt}
\noindent 12: \quad \quad calculate $\bar{E}^{\mathrm{tot},(k)}_{p} = \bar{E}^{\mathrm{inc}}_p - \bar{\bar{G}}\bar{w}^{(k)}_p$\par
\vspace{1pt}
\noindent 13: \quad \quad calculate $\bar{\tau}^{(k)}$ by solving~\eqref{contrast_linear_system}\par
\vspace{1pt}
\noindent 14: \quad \quad $\bar{n}^{(k)} = \mathit{Denoise}(\bar{\tau}^{(k)})$\par
\vspace{1pt}
\noindent 15: \quad \textbf{end for}\par
\vspace{1pt}
\noindent 16: \textbf{end for}\par
\vspace{1pt}
\noindent 17: \textbf{return} $\bar{\tau} = \mathit{Denoise}(\bar{\tau}^{(N_{\mathrm{iter}})})$\par
\vspace{1pt}
\vspace{2pt}\hrule
\end{figure}

\section{Numerical Examples}\label{numerical_examples}

The proposed STO-PNP-CSI is applied to both synthetic and experimental data to assess its performance against other methods. In all examples, the pretrained SCUNet model for denoising grayscale images at noise level $25$ (on the $0$--$255$ scale) is used as the PNP regularizer~\cite{SCUNet}. At each frequency, the number of iterations is set to $N_{\mathrm{iter}}=100$. To evaluate the inversion algorithms, the $\ell_2$-norm relative error between the reconstruction at iteration $k$ and the ground truth is defined as
\begin{equation}
\mathit{err}^{(k)} = \frac{\| \bar{\varepsilon}^{(k)} - \bar{\varepsilon}^{\mathrm{true}} \|_2} {\| \bar{\varepsilon}^{\mathrm{true}} \|_2}\label{relative_error}
\end{equation}
where $\bar{\varepsilon}^{\mathrm{true}}$ is the relative permittivity profile of the ground truth and $\bar{\varepsilon}^{(k)} = \bar{\tau}^{(k)} + 1$ is the reconstructed profile at iteration $k$.

The proposed STO-PNP-CSI is compared with five methods that together isolate the contributions of its two components, stochastic optimization and PNP regularization. CSI is the standard scheme described in Section~\ref{csi}, without stochastic optimization or additional regularization. STO-CSI augments CSI with the stochastic strategy alone, updating a single randomly selected contrast source per iteration. PNP-CSI augments CSI with the PNP regularizer alone, replacing the substep~II update with the SCUNet denoiser while retaining the full per-iteration update over all transmitters. STO-L1-CSI and STO-L2-CSI combine the stochastic strategy with a conventional hand-crafted regularizer in place of the PNP denoiser, using $R(\bar{\tau}) = \|\bar{\tau}\|_1$ and $R(\bar{\tau}) = \|\bar{\tau}\|_2^2$ in~\eqref{sto_pnp_objective_function}, respectively. Comparing these methods against the full STO-PNP-CSI isolates the effect of each component and, in particular, contrasts the PNP denoiser with conventional regularizers.

\subsection{Synthetic Data Inversion}\label{synthetic_data}

\begin{figure}[ht]
\centering
\includegraphics[width=0.45\columnwidth]{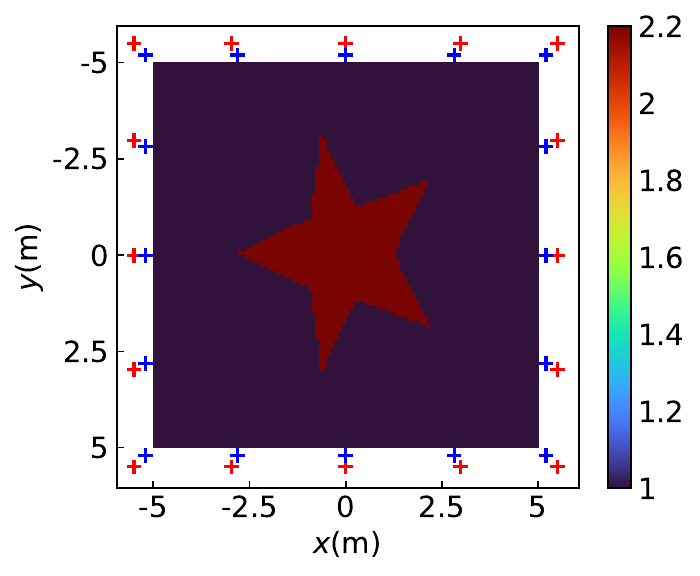}
\caption{System configuration for the synthetic-data examples. The red and blue ``+'' symbols indicate the transmitters and receivers, respectively.}\label{synthetic_configuration}
\end{figure}

All synthetic models share the same system configuration. As shown in Fig.~\ref{synthetic_configuration}, $N_{\mathrm{t}}=16$ transmitters and $N_{\mathrm{r}}=16$ receivers are located around the $10\,\mathrm{m} \times 10\,\mathrm{m}$ investigation domain $S$. The operating frequencies are $f = \{ 100, 200, 300 \}\,\mathrm{MHz}$, so $S$ spans $3.33 \lambda_{0} \times 3.33 \lambda_{0}$ at $100\,\mathrm{MHz}$ and $10 \lambda_{0} \times 10 \lambda_{0}$ at $300\,\mathrm{MHz}$, where $\lambda_{0}$ denotes the wavelength in the background medium. For discretization, $S$ is meshed into $N=100 \times 100$ square cells. Unless stated otherwise, the measured data are obtained by adding Gaussian noise to the scattered fields computed by solving the forward problem with $\bar{\varepsilon}^{\mathrm{true}}$, at a signal-to-noise ratio (SNR) of $30\,\mathrm{dB}$. The penalty weight $\rho$ is empirically set to $\rho=2000$ for PNP-CSI and $\rho=150$ for STO-PNP-CSI. For comparison, the parameters of STO-L1-CSI are set to $\lambda=2$ and $\rho=10$, while $\lambda=10$ is used for STO-L2-CSI.

\subsubsection{Convergence and Consistency}
\begin{sidewaysfigure}[ht]
\centering
\includegraphics[width=\columnwidth]{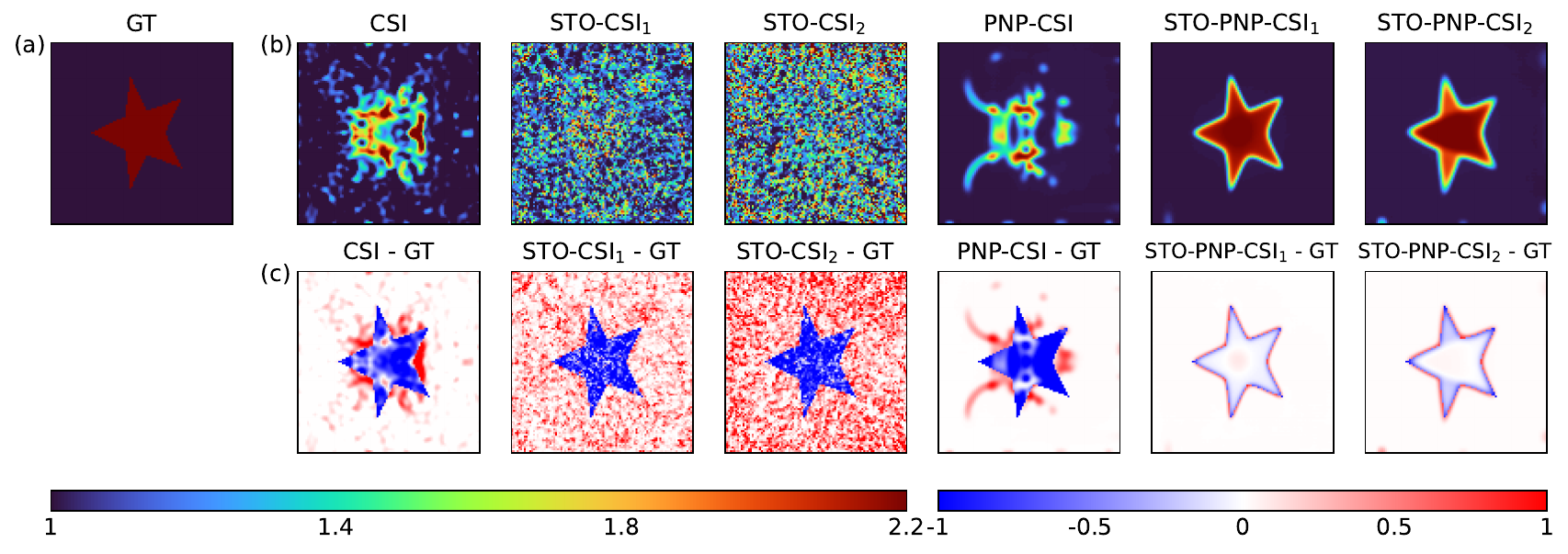}
\caption{Inversion of model I. (a) Ground truth, (b) reconstructions, and (c) their residuals with respect to the ground truth. In (b) and (c), the columns correspond, from left to right, to CSI, STO-CSI\textsubscript{1} (lowest-$\mathit{err}$ run), STO-CSI\textsubscript{2} (highest-$\mathit{err}$ run), PNP-CSI, STO-PNP-CSI\textsubscript{1} (lowest-$\mathit{err}$ run), and STO-PNP-CSI\textsubscript{2} (highest-$\mathit{err}$ run).}\label{distribution_result}
\end{sidewaysfigure}

\begin{figure}[ht]
\centering
\includegraphics[width=0.7\columnwidth]{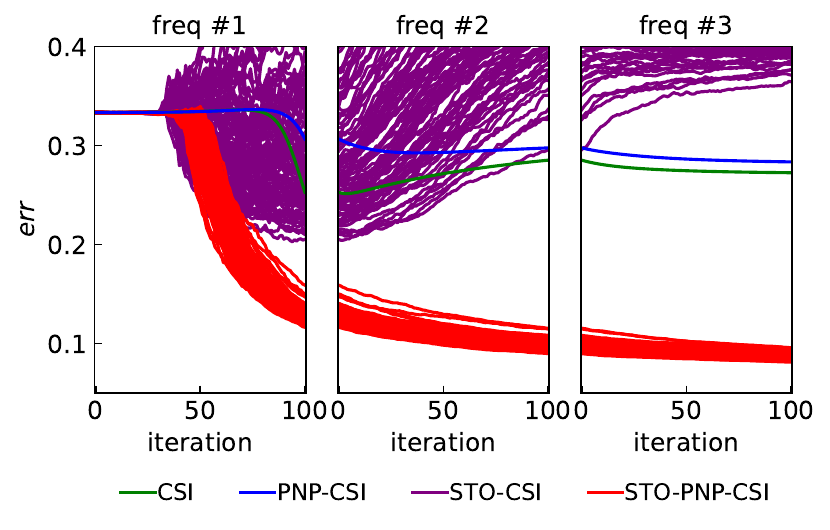}
\caption{The $\mathit{err}$ convergence curves for the inversion of model I.}\label{distribution_convergence}
\end{figure}

Model I is a star-shaped scatterer with a relative permittivity of $2.2$, as shown in Fig.~\ref{distribution_result}(a). Because a fresh transmitter is drawn at random at each iteration, STO-CSI and STO-PNP-CSI can produce different reconstructions from the same data. To characterize this behavior, STO-CSI and STO-PNP-CSI are each executed $100$ times to reconstruct model I. CSI and PNP-CSI are run only once, as they are deterministic and produce identical results across runs. Fig.~\ref{distribution_convergence} shows the $\mathit{err}$ convergence curves obtained for the four methods, and Fig.~\ref{distribution_result}(b) and (c) show the corresponding reconstructions and their residuals with respect to the ground truth. Here, STO-CSI\textsubscript{1} and STO-PNP-CSI\textsubscript{1} denote the reconstructions with the lowest $\mathit{err}$, and STO-CSI\textsubscript{2} and STO-PNP-CSI\textsubscript{2} denote the reconstructions with the highest $\mathit{err}$ among the $100$ runs. From the reconstructions and convergence curves, it can be seen that CSI and PNP-CSI fail to reconstruct the scatterer, with the inversion trapped in local minima. STO-CSI, in turn, fails to converge in any of the $100$ runs, because it lacks a strong regularizer to compensate for the strong ill-posedness introduced by the stochastic updates. Finally, STO-PNP-CSI converges reliably to a much lower $\mathit{err}$ than the other methods across all $100$ runs. Having established this consistency, only the result of a single representative run is reported in each of the following examples.

\subsubsection{Effect of Nonlinearity}
\begin{sidewaysfigure}[ht]
\centering
\includegraphics[width=\textwidth]{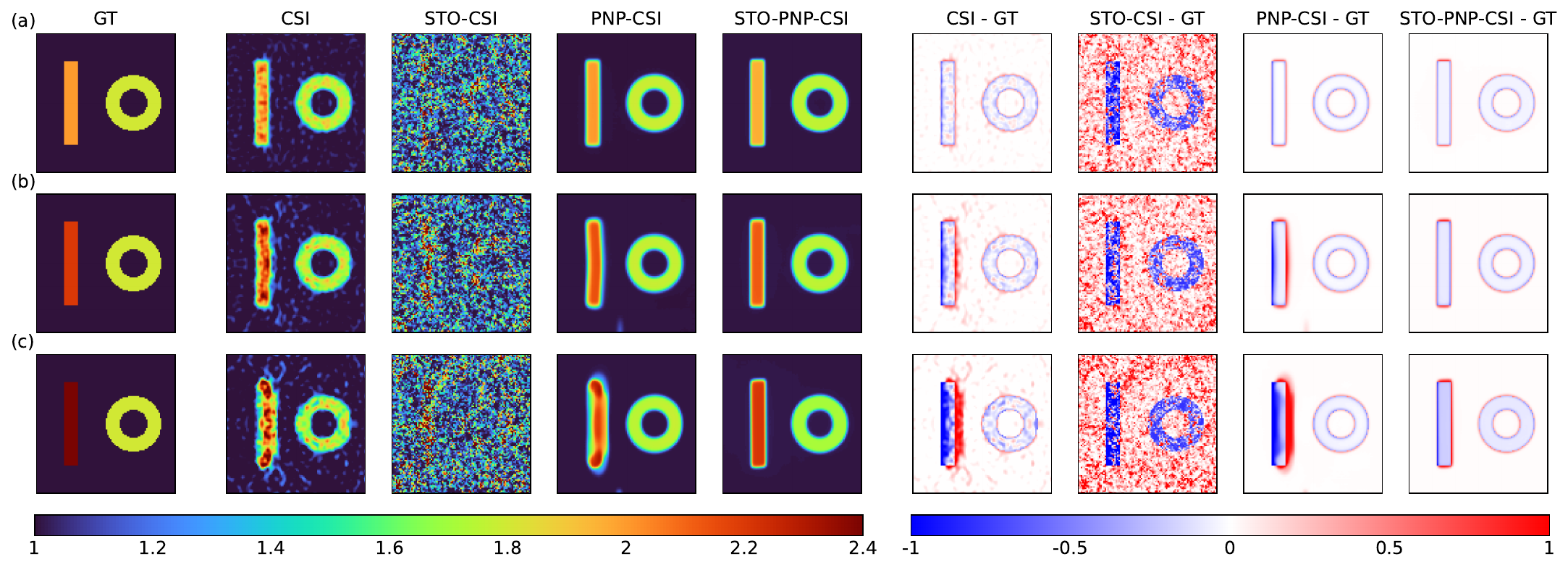}
\caption{Inversion of (a) model II, (b) model III, and (c) model IV. In each of (a)--(c), from left to right are the ground truth, the reconstructions obtained by CSI, STO-CSI, PNP-CSI, and STO-PNP-CSI, and their residuals with respect to the ground truth.}\label{nonlinearity_result}
\end{sidewaysfigure}

In this example, the ability of STO-PNP-CSI to handle strong nonlinearity is examined. As shown in column one of Fig.~\ref{nonlinearity_result}(a)--(c), models II--IV each consist of a common ring-shaped scatterer with a relative permittivity of $1.8$ and a rectangular scatterer with a relative permittivity of $2$, $2.2$, and $2.4$, respectively. Increasing the contrast leads to stronger nonlinearity~\cite{Degree of nonlinearity and a new solution procedure in scalar two-dimensional inverse scattering problems, A new integral equation method to solve highly nonlinear inverse scattering problems}. As in the previous example, CSI, STO-CSI, PNP-CSI, and STO-PNP-CSI are used to reconstruct models II--IV. Columns two to five in Fig.~\ref{nonlinearity_result}(a)--(c) show the corresponding reconstructions, and columns six to nine show their residuals with respect to the ground truth. The CSI reconstructions are heavily corrupted by artifacts. In addition, the reconstructed rectangular scatterer shows a location shift, especially at higher contrast, as seen more clearly in the residuals in column six of Fig.~\ref{nonlinearity_result}(a)--(c). STO-CSI fails to converge for any of the three models, as it lacks sufficient data and regularization. PNP-CSI suppresses the artifacts but cannot correct the location error of the scatterer, as seen in column eight of Fig.~\ref{nonlinearity_result}(a)--(c). 

\begin{figure}[ht]
\centering
\includegraphics[width=0.8\columnwidth]{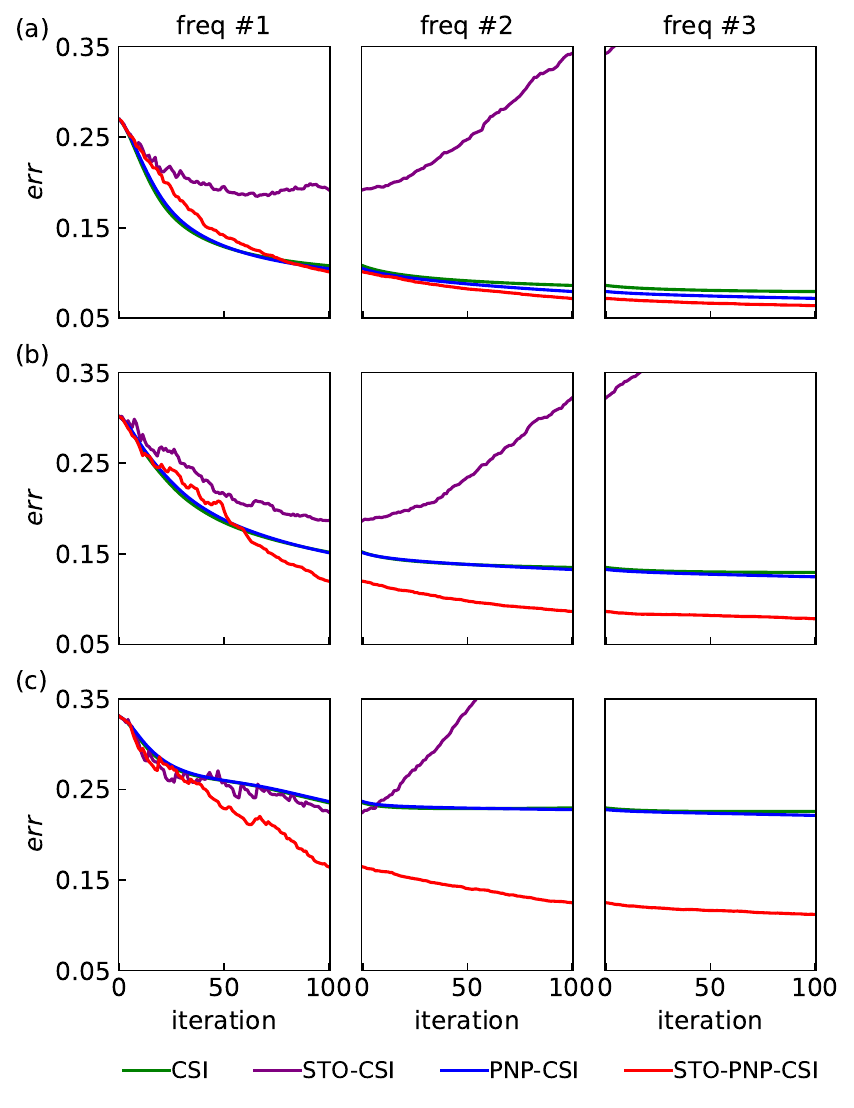}
\caption{The $\mathit{err}$ convergence curves for the inversion of (a) model II, (b) model III, and (c) model IV.}\label{nonlinearity_convergence}
\end{figure}

Finally, STO-PNP-CSI combines the strengths of STO-CSI and PNP-CSI: the stochastic updates can help the inversion escape poor local minima (here, correcting the position of the rectangular scatterer), while the PNP regularizer compensates for the loss of information and stabilizes the inversion (here, promoting the convergence of the stochastic updates and suppressing artifacts). Consequently, STO-PNP-CSI achieves the most accurate localization of the rectangular scatterer and the best reconstruction of the relative permittivity profile. Fig.~\ref{nonlinearity_convergence} shows the $\mathit{err}$ convergence curves of all methods. The advantage of STO-PNP-CSI becomes more pronounced as the nonlinearity increases: it converges to the lowest $\mathit{err}$ of the four methods, and the gap widens from model II to model IV.

\subsubsection{Effect of Noise}
\begin{sidewaysfigure}[ht]
\centering
\includegraphics[width=\textwidth]{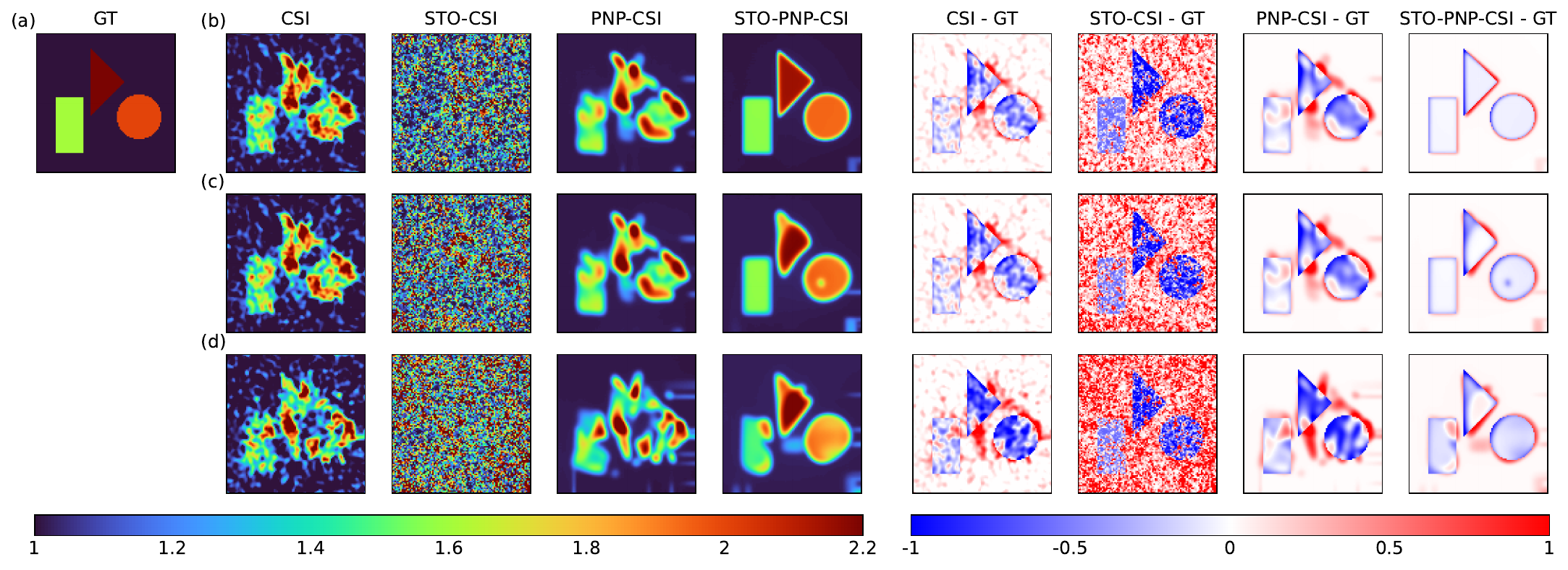}
\caption{Inversion of model V. (a) Ground truth. (b)--(d) Reconstructions and residuals for data with SNRs of (b) $10\,\mathrm{dB}$, (c) $20\,\mathrm{dB}$, and (d) $30\,\mathrm{dB}$. In each of (b)--(d), from left to right are the reconstructions obtained by CSI, STO-CSI, PNP-CSI, and STO-PNP-CSI, followed by their residuals with respect to the ground truth.}\label{noise_result}
\end{sidewaysfigure}

\begin{figure}[ht]
\centering
\includegraphics[width=0.8\columnwidth]{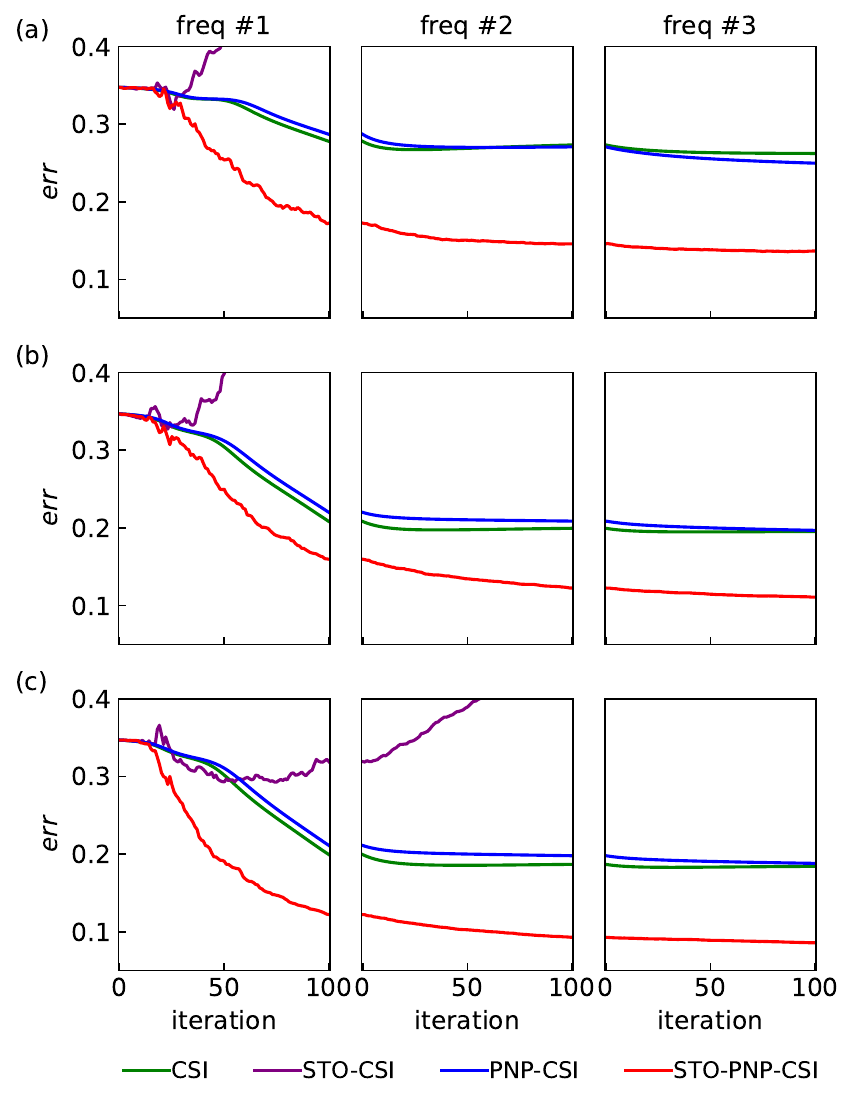}
\caption{The $\mathit{err}$ convergence curves for the inversion of model V, for data with SNRs of (a) $10\,\mathrm{dB}$, (b) $20\,\mathrm{dB}$, and (c) $30\,\mathrm{dB}$.}\label{noise_convergence}
\end{figure}

This example examines the robustness of STO-PNP-CSI to measurement noise. As shown in Fig.~\ref{noise_result}(a), model V consists of triangular, circular, and rectangular scatterers with relative permittivities of $2.2$, $2$, and $1.6$, respectively. The measured data are generated at SNRs of $10$, $20$, and $30\,\mathrm{dB}$. The reconstructions obtained by CSI, STO-CSI, PNP-CSI, and STO-PNP-CSI and their residuals with respect to the ground truth at $10$, $20$, and $30\,\mathrm{dB}$ noise levels are shown in Fig.~\ref{noise_result}(b), (c), and (d), respectively. STO-PNP-CSI produces the most accurate reconstructions among the four methods at every noise level, and its accuracy degrades only gradually as the SNR decreases. This indicates that STO-PNP-CSI is robust to noise, which is also corroborated by the $\mathit{err}$ convergence curves in Fig.~\ref{noise_convergence}. Table~\ref{model_v_metrics} reports the value of $\mathit{err}$ at the last iteration and the computation time of each method. For CSI and PNP-CSI, which solve for the contrast source vectors of all $N_{\mathrm{t}}$ transmitters at every iteration, parallel computing is used to accelerate the computation. Therefore, both the total core time and the wall-clock time are provided for a fair comparison. As the table shows, STO-PNP-CSI achieves the lowest $\mathit{err}$ while requiring the least computation---its core time is more than an order of magnitude below that of CSI and PNP-CSI, and its wall-clock time is also the lowest.

\subsubsection{Comparison of Regularizers}
\begin{sidewaysfigure}[ht]
\centering
\includegraphics[width=\textwidth]{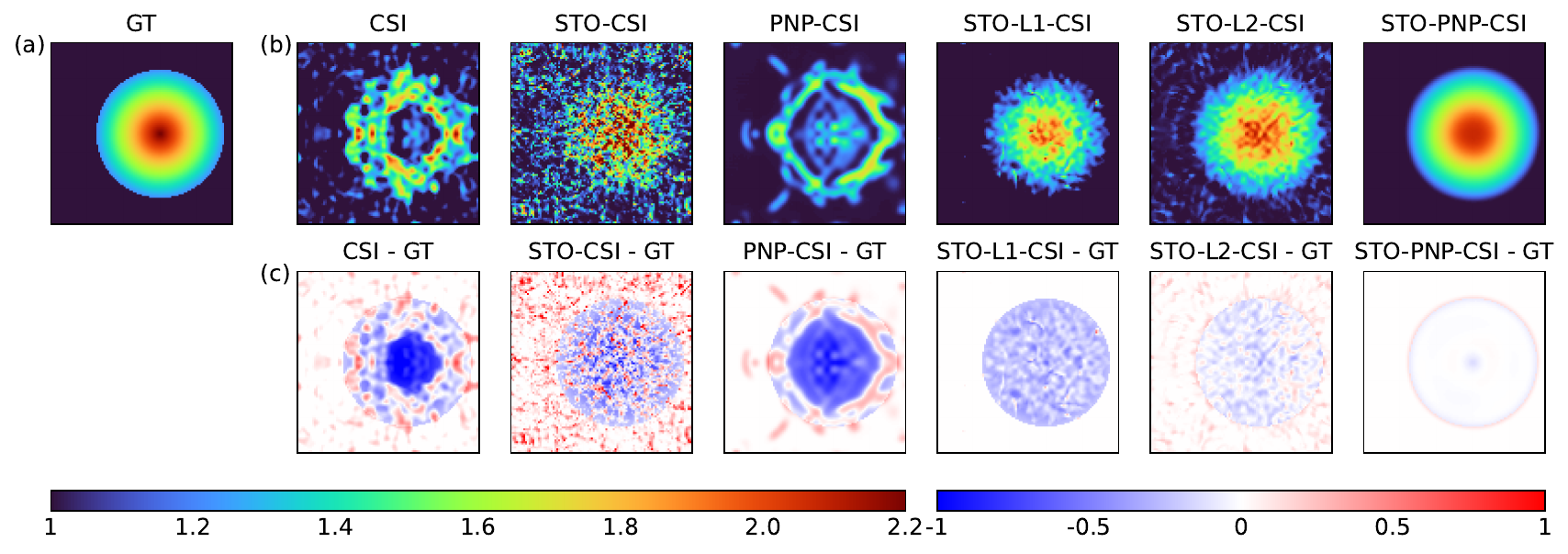}
\caption{Inversion of model VI. (a) Ground truth, (b) reconstructions, and (c) their residuals with respect to the ground truth. In (b) and (c), the columns correspond, from left to right, to CSI, STO-CSI, PNP-CSI, STO-L1-CSI, STO-L2-CSI, and STO-PNP-CSI.}\label{regularization_result}
\end{sidewaysfigure}

\begin{figure}[ht]
\centering
\includegraphics[width=0.8\columnwidth]{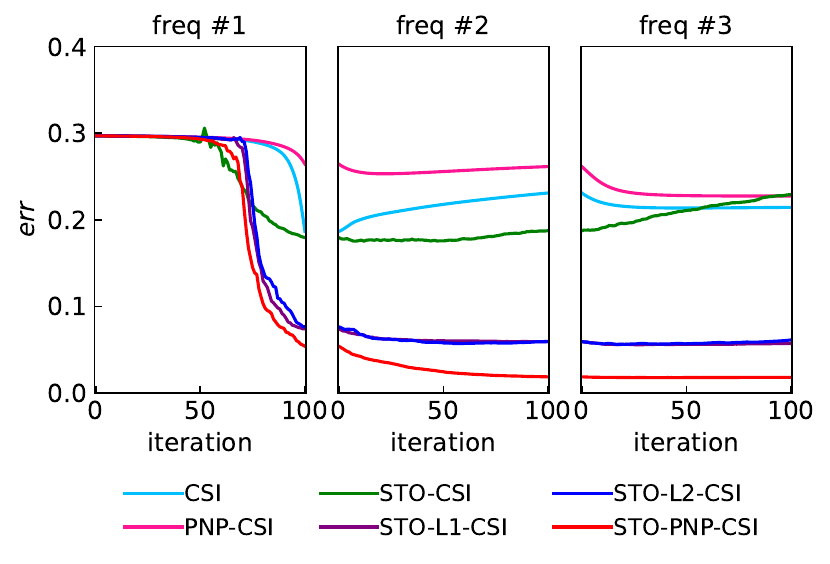}
\caption{The $\mathit{err}$ convergence curves for the inversion of model VI.}\label{regularization_convergence}
\end{figure}

As noted in Section~\ref{formulation} and shown in the previous examples, STO-CSI fails to converge without a strong regularizer. This example compares the performance of STO-CSI incorporating different regularization methods. As shown in Fig.~\ref{regularization_result}(a), model VI is a circular scatterer with an inhomogeneous relative permittivity distribution that peaks at $2.2$ at its center. In addition to CSI, STO-CSI, PNP-CSI, and STO-PNP-CSI, two further variants are included for comparison: STO-L1-CSI, which uses $\ell_1$-norm regularization ($R(\bar{\tau}) = \| \bar{\tau} \|_1$), and STO-L2-CSI, which uses $\ell_2$-norm regularization ($R(\bar{\tau}) = \| \bar{\tau} \|_2^2$). The reconstructions are shown in Fig.~\ref{regularization_result}(b), and the corresponding residuals with respect to the ground truth in Fig.~\ref{regularization_result}(c). Without the stochastic updates, CSI and PNP-CSI fail to produce reasonable reconstructions. When the stochastic updates are used alone (STO-CSI), the reconstruction is heavily polluted. The results improve substantially only when the stochastic updates are combined with regularization, as in STO-L1-CSI, STO-L2-CSI, and STO-PNP-CSI. Among these, the conventional $\ell_1$- and $\ell_2$-norm regularizers cannot fully suppress the artifacts introduced by the stochastic updates. STO-PNP-CSI reconstructs the inhomogeneous profile most accurately, because it employs a neural-network denoiser as the regularizer. This is consistent with the $\mathit{err}$ convergence curves in Fig.~\ref{regularization_convergence}. Overall, this example indicates that the stochastic scheme requires strong regularization to be effective, and that the PNP regularizer is the most effective among the regularizers tested. 

\subsection{Experimental Data Inversion}\label{experimental_data}

\begin{sidewaysfigure}[ht]
\centering
\includegraphics[width=0.9\columnwidth]{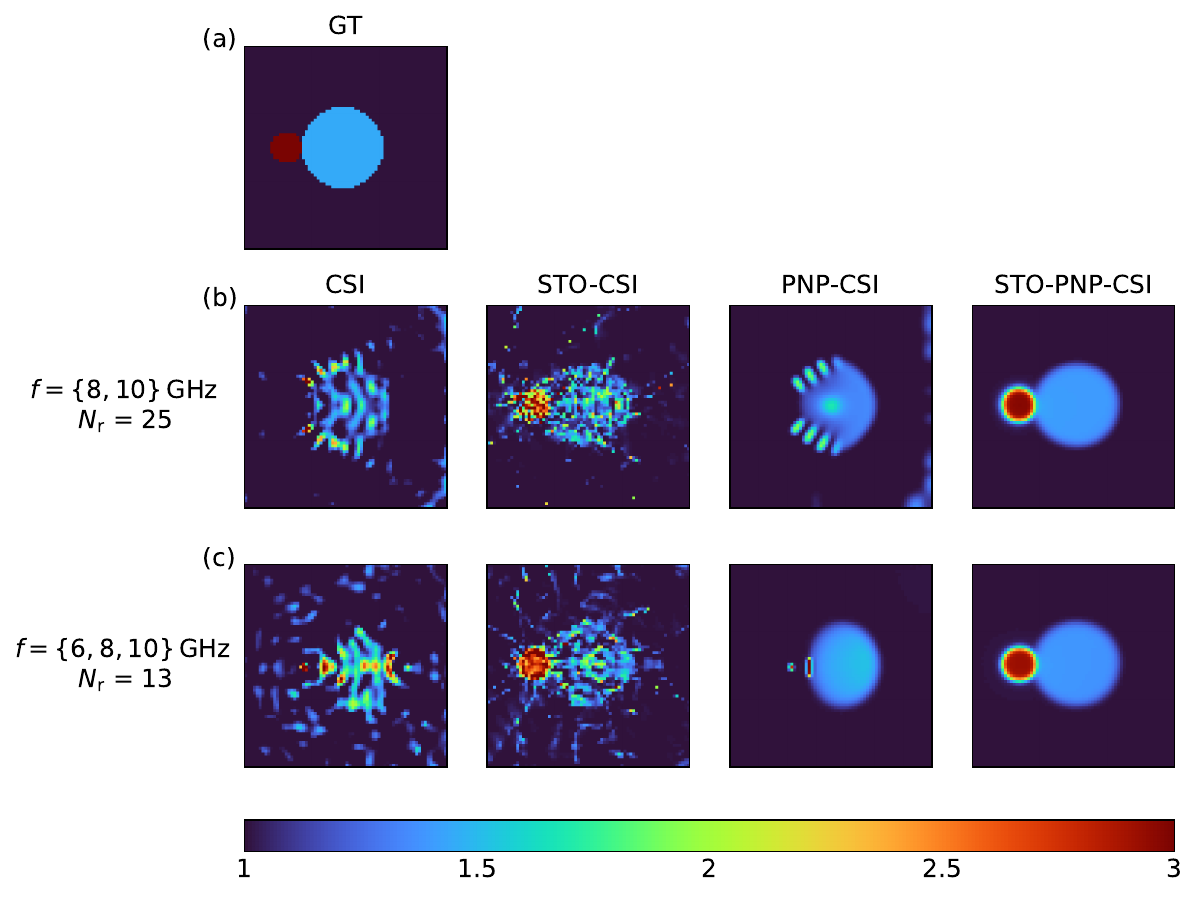}
\caption{Inversion of the ``FoamDielExtTM'' target. (a) Ground truth, and reconstructions for (b) $f = \{8, 10\}\,\mathrm{GHz}$, $N_{\mathrm{r}} = 25$ and (c) $f = \{6, 8, 10\}\,\mathrm{GHz}$, $N_{\mathrm{r}} = 13$. In (b) and (c), the columns correspond, from left to right, to CSI, STO-CSI, PNP-CSI, and STO-PNP-CSI.}\label{fresnel_results}
\end{sidewaysfigure}

In this subsection, STO-PNP-CSI is applied to experimental data from the Fresnel database~\cite{Free space experimental scattering database continuation: Experimental set-up and measurement precision}. As shown in Fig.~\ref{fresnel_results}(a), the ``FoamDielExtTM'' target, located in a $0.2 \, \mathrm{m} \times 0.2 \, \mathrm{m}$ investigation domain, consists of two cylindrical scatterers: a foam cylinder with a relative permittivity of $1.45$ and a radius of $0.04\,\mathrm{m}$, and a smaller plastic cylinder with a relative permittivity of $3$ and a radius of $0.0155\,\mathrm{m}$. The cylinders are long along the $z$-axis, so the problem is treated as a 2-D TM inversion. A transmitting and a receiving horn antenna are placed on a circle of radius $1.67\,\mathrm{m}$. The transmitter illuminates the target from $N_{\mathrm{t}}=8$ locations spaced $45^\circ$ apart, one at a time. For each illumination, the receiver measures the scattered field at $241$ locations spaced $1^\circ$ apart, covering the $240^\circ$ arc facing away from the transmitter. The system operates at $9$ frequencies from $2$ to $10 \, \mathrm{GHz}$. Because the receiver locations are defined relative to each transmitter, the receiver set differs from transmitter to transmitter. Accordingly, the integrated Green function matrix $\bar{\bar{H}}$ is transmitter-dependent in this experiment, unlike in the synthetic examples where it is common to all transmitters. To make the inversion more challenging, the measured data are down-sampled. Specifically, two system configurations are considered. In the first configuration, the number of receiver locations is reduced to $N_{\mathrm{r}}=25$ (spaced $10^\circ$ apart) and the frequencies to $f = \{8, 10\} \, \mathrm{GHz}$. In the second configuration, the number of receiver locations is reduced to $N_{\mathrm{r}}=13$ (spaced $20^\circ$ apart) and the frequencies to $f = \{6, 8, 10\} \, \mathrm{GHz}$.   For both configurations, the penalty weight $\rho$ is empirically set to $\rho = 0.02$ for PNP-CSI and $\rho = 0.0015$ for STO-PNP-CSI. $S$ is meshed into $N = 70 \times 70$ cells. CSI, STO-CSI, PNP-CSI, and STO-PNP-CSI are then applied.

Fig.~\ref{fresnel_results}(b) and (c) show the reconstructions for the two configurations. Under the strong nonlinearity and ill-posedness of this problem, CSI does not produce an acceptable reconstruction in either configuration. Adding either the stochastic updates (STO-CSI) or the PNP regularizer (PNP-CSI) lowers $\mathit{err}$ relative to that of CSI, but neither alone recovers the target: STO-CSI is heavily corrupted by noise, and PNP-CSI fails to recover the small high-contrast scatterer. STO-PNP-CSI accurately reconstructs both scatterers in both configurations. Table~\ref{fresnel_metrics} reports the value of $\mathit{err}$ at the last iteration and the computation time of each method for the two configurations. STO-PNP-CSI attains the lowest $\mathit{err}$ in both cases. For CSI and PNP-CSI, the per-transmitter solves are parallelized, so their wall-clock time is much lower than their core time. CSI requires the lowest wall-clock time, whereas STO-PNP-CSI requires the lowest core time. Overall, combining the stochastic updates with the PNP regularizer yields accurate reconstructions under these challenging conditions while keeping the total computational cost low.

\section{Conclusion}\label{conclusion}

An STO-PNP-CSI scheme is developed to accelerate CSI and improve its robustness for highly nonlinear inverse scattering problems. Building on the well-established CSI framework, STO and PNP regularization are introduced, together with a frequency-hopping strategy for multi-frequency measurements. By updating only one randomly selected contrast source vector at each iteration, STO greatly reduces the per-iteration cost and helps the inversion escape poor local minima. To counter the ill-posedness introduced by loss of information due to the stochastic update scheme, PNP regularization uses a neural-network denoiser as an implicit regularizer, supplying a strong prior that stabilizes the reconstruction. The proposed STO-PNP-CSI is evaluated on both synthetic and experimental data. The results show that it yields accurate reconstructions at substantially lower computational cost than CSI, including under strong nonlinearity and measurement noise.

\clearpage\newpage

\section*{Figures}

\newpage

\newpage

\newpage

\newpage

\newpage

\newpage

\newpage

\newpage

\newpage

\newpage

\clearpage\newpage

\section*{Tables}
\begin{table}[ht]
\caption{Quantitative Analysis of Model V}\label{model_v_metrics}
\centering
\renewcommand{\arraystretch}{1.2}
\begin{tabular}{ccrrr}
SNR (dB) & Method & $\mathit{err}$ & \makecell{Core \\ time (s)} & \makecell{Wall-clock \\ time (s)} \\
\hline
\multirow{4}{*}{10} & CSI & 0.2623 & 8306.72 & 640.50 \\
 & STO-CSI & 0.7202 & 2176.57 & 2173.36 \\
 & PNP-CSI & 0.2497 & 8650.31 & 763.91 \\
 & STO-PNP-CSI & \textbf{0.1367} & \textbf{569.85} & \textbf{565.45} \\
\hline
\multirow{4}{*}{20} & CSI & 0.1954 & 8376.86 & 646.92 \\
 & STO-CSI & 0.5762 & 2627.18 & 2623.74 \\
 & PNP-CSI & 0.1969 & 8500.44 & 752.17 \\
 & STO-PNP-CSI & \textbf{0.1109} & \textbf{577.27} & \textbf{573.67} \\
\hline
\multirow{4}{*}{30} & CSI & 0.1842 & 8382.12 & 648.38 \\
 & STO-CSI & 0.4640 & 2122.60 & 2119.89 \\
 & PNP-CSI & 0.1880 & 8495.27 & 755.21 \\
 & STO-PNP-CSI & \textbf{0.0858} & \textbf{557.66} & \textbf{552.85} \\
\end{tabular}
\end{table}

\newpage
\begin{table}[ht]
\caption{Quantitative Analysis of the ``FoamDielExtTM'' Target}\label{fresnel_metrics}
\centering
\renewcommand{\arraystretch}{1.2}
\begin{tabular}{ccrrr}
\makecell{System \\ configuration} & Method & $\mathit{err}$ & \makecell{Core \\ time (s)} & \makecell{Wall-clock \\ time (s)} \\
\hline
\multirow{4}{*}{\makecell{$f = \{8, 10\}\,\mathrm{GHz}$ \\ $N_{\mathrm{r}} = 25$}} & CSI & 0.2878 & 496.20 & \textbf{106.75} \\
 & STO-CSI & 0.2181 & 161.72 & 159.40 \\
 & PNP-CSI & 0.2590 & 450.85 & 145.53 \\
 & STO-PNP-CSI & \textbf{0.0978} & \textbf{140.17} & 137.94 \\
\hline
\multirow{4}{*}{\makecell{$f = \{6, 8, 10\}\,\mathrm{GHz}$ \\ $N_{\mathrm{r}} = 13$}} & CSI & 0.3003 & 615.96 & \textbf{108.91} \\
 & STO-CSI & 0.1763 & 223.76 & 220.82 \\
 & PNP-CSI & 0.2580 & 636.06 & 180.62 \\
 & STO-PNP-CSI & \textbf{0.1009} & \textbf{190.83} & 187.64 \\
\end{tabular}
\end{table}

\end{document}